# Short-term synaptic plasticity in the deterministic Tsodyks–Markram model leads to unpredictable network dynamics

Jesus M. Cortes[a,b,1], Mathieu Desroches[c,1], Serafim Rodrigues[d,1], Romain Veltz[e], Miguel A. Muñoz[f], and Terrence J. Sejnowski[e,g,h,2]

[a]Ikerbasque, Basque Foundation for Science, 48011 Bilbao, Spain; [b]Biocruces Health Research Institute, Hospital Universitario de Cruces, 48903 Barakaldo, Spain; [c]Institut National de Recherche en Informatique et en Automatique Paris-Rocquencourt Centre, 78153 Le Chesnay Cedex, France; [d]School of Computing and Mathematics, University of Plymouth, Plymouth PL4 8AA, United Kingdom; [e]The Computational Neurobiology Laboratory and [g]Howard Hughes Medical Institute, Salk Institute, La Jolla, CA 92037; [f]Facultad de Ciencias, Departamento de Electromagnetismo y Física de la Materia and Instituto Carlos I de Física Teórica y Computacional, Universidad de Granada, 18012 Granada, Spain; and [h]Division of Biological Science, University of California, San Diego, La Jolla, CA 92093

Contributed by Terrence J. Sejnowski, August 28, 2013 (sent for review April 13, 2013)

**Short-term synaptic plasticity strongly affects the neural dynamics of cortical networks. The Tsodyks and Markram (TM) model for short-term synaptic plasticity accurately accounts for a wide range of physiological responses at different types of cortical synapses. Here, we report a route to chaotic behavior via a Shilnikov homoclinic bifurcation that dynamically organizes some of the responses in the TM model. In particular, the presence of such a homoclinic bifurcation strongly affects the shape of the trajectories in the phase space and induces highly irregular transient dynamics; indeed, in the vicinity of the Shilnikov homoclinic bifurcation, the number of population spikes and their precise timing are unpredictable and highly sensitive to the initial conditions. Such an irregular deterministic dynamics has its counterpart in stochastic/network versions of the TM model: The existence of the Shilnikov homoclinic bifurcation generates complex and irregular spiking patterns and—acting as a sort of springboard—facilitates transitions between the down-state and unstable periodic orbits. The interplay between the (deterministic) homoclinic bifurcation and stochastic effects may give rise to some of the complex dynamics observed in neural systems.**

Short-term synaptic plasticity (STSP) is a temporal increase or decrease of synaptic strength in response to use, modulating the efficacy of synaptic transmission on timescales from a few milliseconds to several minutes (1–3). Short-term variations of the synaptic efficacy directly affect the timing and integration of presynaptic inputs to postsynaptic neurons, thereby affecting network-level dynamics and influencing brain function. Synaptic depression, an activity-regulatory mechanism that performs gain control (4), reduces cortical responses to strong stimuli (5). Recurrent cortical circuits in the presence of synaptic depression work as nonlinear adaptive filters in the visual system, with long response latencies for low-contrast stimuli and short latencies for high-contrast stimuli (6). Synaptic depression affects the responses of individual neurons and also the correlations of neural populations and the population accuracy for coding signals (7). Finally, working memory can be implemented through STSP using the dynamics of synaptic facilitation (8). Given this broad spectrum of disparate functions, it is not surprising that different cortical synapses in different cortical areas exhibit vastly different STSP dynamics (9).

The Tsodyks and Markram (TM) model describes the dynamics of STSP under general conditions in an elegant and parsimonious way (1, 2). Indeed, the TM model includes several parameters that can be identified with biophysical variables (such as the time constant for refilling the readily releasable pool of vesicles) that can be regulated and account remarkably well for the highly heterogeneous dynamics in areas such as the visual and prefrontal cortices (see table 1 in ref. 10). Moreover, the TM model can explain as well the transitions between up and down cortical states (11–13).

Here, we report the existence of highly irregular and chaotic-like dynamics in the TM model. Chaotic dynamics has been already suggested as a possible mechanism to explain the irregular dynamics observed in cortical activity (14), as reflected, for example, in trial-to-trial variability of neuron responses observed after multiple repetitions of the same stimulus (15, 16), and the erratic and complex transients occurring in local field potentials (17). We identified a form of chaos in the TM model, called Shilnikov chaos, which induces highly irregular transient dynamics and large sensitivity to initial conditions, even for the case of the Shilnikov chaos having an associated attractor that is unstable, rather than a stable one (18–21). The existence of Shilnikov chaos cannot, in principle, account for all of the observed complexity of cortical dynamics, but it may account for some irregularity in the overall network activity and, in particular, for facilitating transitions between large-scale brain states.

## Model

We consider the continuous time TM model (8) as sketched in Fig. 1; the corresponding differential equations are detailed in *Materials and Methods*. The model describes the deterministic behavior of a population of identical excitatory neurons (in blue

### Significance

**Short-term synaptic plasticity contributes to the balance and regulation of brain networks from milliseconds to several minutes. In this paper we report the existence of a route to chaos in the Tsodyks and Markram model of short-term synaptic plasticity. The chaotic region corresponds to what in mathematics is called Shilnikov chaos, an unstable manifold that strongly modifies the shape of trajectories and induces highly irregular transient dynamics, even in the absence of noise. The interplay between the Shilnikov chaos and stochastic effects may give rise to some of the complex dynamics observed in neural systems such as transitions between up and down states.**

Author contributions: J.M.C., M.D., S.R., and T.J.S. designed research; J.M.C., M.D., S.R., R.V., M.A.M., and T.J.S. performed research; J.M.C., M.D., S.R., R.V., M.A.M., and T.J.S. analyzed data; and J.M.C., M.D., S.R., R.V., M.A.M., and T.J.S. wrote the paper.

The authors declare no conflict of interest.

Freely available online through the PNAS open access option.

[1]J.M.C., M.D., and S.R. contributed equally to this work.

[2]To whom correspondence should be addressed. E-mail: terry@salk.edu.

in Fig. 1*A*) described by three state variables: the average excitatory network activity at any given time, $E(t)$ (in hertz) and two other variables, $x(t)$ and $u(t)$, that model synaptic depression and facilitation, respectively. Each neuron receives global inhibition (in red in Fig. 1*A*) represented by the constant $I_0$.

In the TM model, presynaptic resources are finite and each one is represented by two possible states: available to be released, inducing presynaptic activation (green circles in Fig. 1*B*), and nonavailable for release (orange circles in Fig. 1*B*). The overall fraction of available neurotransmitters is $x(t)$, whereas $1-x(t)$ is the fraction of nonavailable ones. Network activity, $E(t)>0$, consumes resources; the consumption rate from state $x(t)$ to $1-x(t)$ is given by $u(t)E(t)$ (discussed below), which leads to short-term synaptic depression. The transition from non-available to available states (i.e., refilling dynamics) is occurring at a rate $\tau_D^{-1}$, where $\tau_D$ represents the spontaneous recovery time from the depressed state.

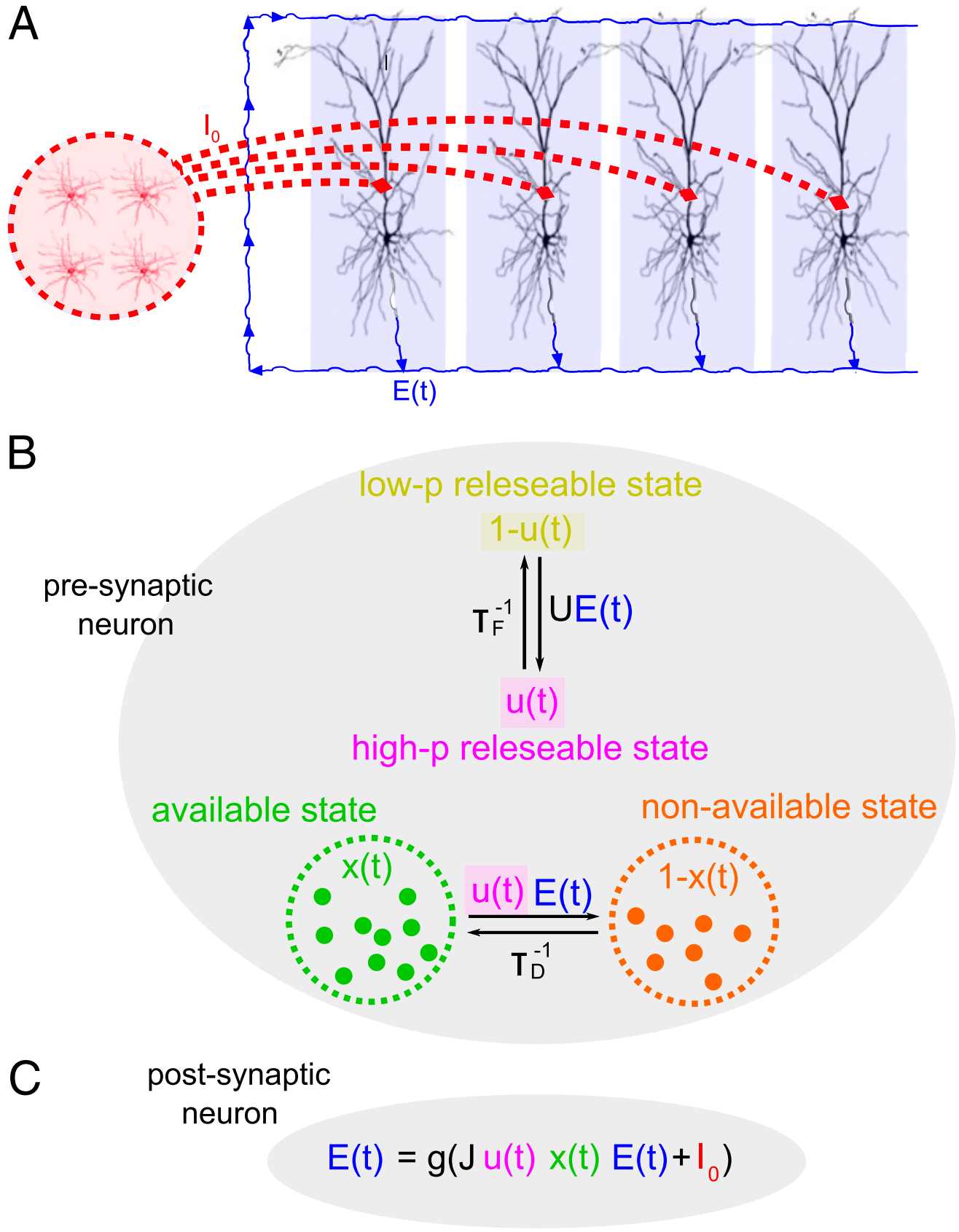

**Fig. 1.** Model of the STSP dynamics. (*A*) Cartoon of a neural network; a pool of identical excitatory pyramidal neurons (blue) are recurrently connected and receive global inhibition from a pool of identical interneurons (red). $E(t)$ is the average rate in hertz for the excitatory pool and $I_0$ the average rate in hertz for inhibitory neurons (fixed parameter). Excitatory connections are described by the TM model for STSP. (*B*) Sketch of TM dynamics. At the presynaptic site (large gray oval) resources are represented by small circles and could be in two possible states: available to be released [green, $x(t)$] and nonavailable [orange, $1-x(t)$]. The transition rates between these states are given by $u(t)E(t)$ and $\tau_D^{-1}$. The variable $u(t)$ (in pink) models variations in release probability, affecting neurotransmitters that have a high probability of releasing $u(t)$ and to those with a low release probability [$1-u(t)$, in yellow]. The transition rates between the two states are given by $\tau_F^{-1}$ and $U\,E(t)$, with $U$ the baseline decay for $u(t)$. (*C*) Equilibrium solution for the average activity in the excitatory population. The total input arriving to the postsynaptic neuron (small gray oval), given by $Ju(t)x(t)E(t)+I_0$ with $J$ a synaptic-strength parameter, is rectified and amplified through the gain function $g$.

The facilitation variable $u(t)$ encodes variations in the release probability of available neurotransmitters, owing mainly to the influx of calcium ions into the presynaptic terminal. Thus, neurotransmitters are represented by two additional states: a fraction $1-u(t)$ (yellow in Fig. 1*B*) with a low probability to be released (the low-p releasable state) and a complementary fraction $u(t)$ in the high-p releasable state (pink in Fig. 1*B*). The transition rate from low-p to high-p releasable state is given by $U\,E(t)$, with $U$ representing the baseline level of $u(t)$, whereas the reverse transition occurs spontaneously at a rate $\tau_F^{-1}$, where $\tau_F$ is the time constant modeling facilitation.

The total input arriving at the postsynaptic terminal is given by $Ju(t)x(t)E(t)+I_0$ (where $J$ is the strength of recurrent connections in the neuronal excitatory population), which is then rectified and amplified via the gain function $g(z)=\alpha\log(1+e^{z/\alpha})$ (ref. 8 and *Materials and Methods*). Parameter values and time constants are summarized in Table S1.

## Results

**Bifurcation Diagram of the TM Model.** The TM model is known to exhibit a wide range of dynamical regimes, such as bistable electrical activity, mimicking cortical up and down states, and oscillatory up states (see figure S1 in ref. 8). Here, we further explored the TM phase space using a numerical software tool, XPPAUT (22), for bifurcation analysis. This software enabled us to systematically track the model's trajectories and stability of equilibrium points and limit cycles while continuously varying system parameters. In addition, XPPAUT allowed us to reconstruct the full bifurcation diagram summarizing all possible states. We identified stable attractor solutions as well as unstable solutions beyond the reach of standard brute-force search. The resulting bifurcation diagram, shown in Fig. 2, illustrates all possible transitions from steady-state activity to oscillatory and complex oscillations under continuous variations of the external inhibition parameter $I_0$. Solid lines correspond to stable solutions and dashed lines to unstable ones. There are three classes of network stable states, which had already been described (8): A down-state fixed point (blue), an up-state fixed point (black), and a periodic up state (a limit cycle whose maximum and minimum amplitudes are the lines marked in red and green, respectively). Further continuing, these states become unstable and we subsequently have detected a Shilnikov homoclinic bifurcation point (orange), which had not previously been identified. Although the trajectories associated with the Shilnikov homoclinic bifurcation are unstable, its presence organizes the observable network states, as we will show below.

**States and Transitions in the TM Model.** For very large inhibition currents, $|I_0|$, the only stable state is the down state. In contrast, for relatively small inhibitory inputs, the system settles to a fixed point with nonvanishing constant activity, called an up state; by progressively increasing $|I_0|$ a saddle-node of a periodic bifurcation leads to a stable oscillatory up state, which coexists with the up state and eventually loses its stability, giving rise to a branch of unstable periodic states. This solution branch terminates at a Shilnikov homoclinic bifurcation (Fig. 2 gives a full description of the different bifurcations).

Note that there is another subcritical Hopf bifurcation at $I_0=-1.84$; the branch of unstable limit cycles emanating from that Hopf bifurcation terminates at a homoclinic bifurcation at $I_0\approx-1.82$, which is not of Shilnikov type. This branch has no impact on the dynamics of the system.

**Properties of the Shilnikov Homoclinic Bifurcation.** The existence of a Shilnikov homoclinic bifurcation in STSP dynamics unveils hidden complexity in the TM model. Indeed, as illustrated in Fig. 3*A*, the period $T$ of the oscillations of the unstable branch of solutions grows without bound as the homoclinic bifurcation is approached, becoming infinitely large right at the bifurcation point; this infinitely large period can be a first indication of chaotic-like dynamics (23).

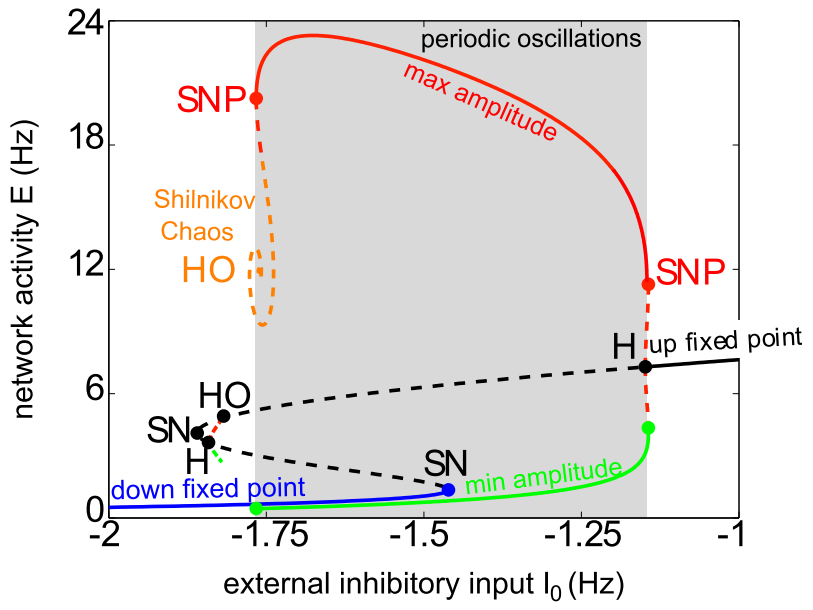

**Fig. 2.** Bifurcation diagram of the STSP dynamics. Numerical continuation of $E$ when varying parameter $I_0$ for $J = 3.07$. Changes in solutions stability occurred at bifurcation points: saddle node at (SN) $I_0 = -1.46$ and $I_0 = -1.86$, subcritical Hopf (H) at $I_0 = -1.15$, and saddle node of a periodic solution (SNP) at $I_0 = -1.14$ and $I_0 = -1.77$. Chaos emerged near the homoclinic bifurcation (HO) at $I_0 \approx -1.76$. Different classes of states (fixed points) describing the network dynamics are depicted with different colors: down-state fixed point (blue), up-state fixed point (black), up-periodic state (red for the maximum oscillation amplitude and green for the minimum), and up-chaotic state (orange). Solid lines refer to stable solutions and dashed lines to unstable solutions. Notice that there is coexistence between different network states, so that for a fixed external stimulus either a small parameter variation or simply noise can produce transitions between the different network states that coexist along the same vertical line in the bifurcation diagram. The SN, H, and SNP bifurcations were reported in ref. 8, but the Shilnikov homoclinic bifurcation HO, which indicates chaos, is new here. Note that there is another subcritical Hopf (H) at $I_0 = -1.84$, creating a branch of unstable limit cycles that terminate at a homoclinic bifurcation (HO) at $I_0 \approx 1.82$. This HO bifurcation does not produces Shilnikov chaos, so it has no impact on the overall dynamics.

As shown in Fig. 3*B* the time series (plotted as a function of $t/T$), along the previously identified wiggly branch, changes its time profile as the homoclinic bifurcation is approached. Far from the homoclinic bifurcation (yellow circle 1 in Fig. 3*A*), the solutions are harmonic oscillations, but as the period of oscillation increases (green circle 2) the solutions exhibit a spiking behavior followed by a slow oscillatory relaxation to a nominal value; this slow return to a quasi-rest state lasts longer and longer as the homoclinic bifurcation is approached. Eventually, at the homoclinic bifurcation the network activity occurs in the form of infinitely separated spikes (purple circle 3), and the period of oscillations increases without bound as the homoclinic orbit is approached (further details in *Materials and Methods*).

An approximate (computationally tractable) solution of a homoclinic orbit is shown in Fig. 3*C*; observe that the orbit is stable in the $(u,x)$-plane and spirals toward a saddle equilibrium fixed point where the stable and unstable manifolds intersect; the orbit then shoots out in the unstable direction, with $E$ increasing in a "spike-like" way, before spiraling back to the saddle-equilibria. In the neighborhood of the homoclinic bifurcation trajectories are similar to this orbit, but eventually after some number of cycles around the attractor, they flow away along the unstable direction. Thus, $E(t)$ exhibits transient population spikes, the number of spikes, their amplitude, and their exact locations being extremely sensitive to initial conditions. This is illustrated in Fig. 4, where different traces of activity corresponding to slightly different initial conditions are shown. The irregular spike-like bursting has an oscillatory character as the system eventually settles to the stable down state. These orbits corresponds to solutions very close to the homoclinic orbit, which repeat themselves with relatively small variance; despite this regularity, the return periods of these orbits widely fluctuate with weak correlation between successive returns (24); these are typical features of homoclinic chaos. In conclusion, the existence of a homoclinic bifurcation profoundly affects the transient dynamics in a broad neighborhood around it, leading to irregular dynamics that directly affects the transient responses of the TM model.

A mathematical test for whether the dynamics around the attractor is actually chaotic is to evaluate the so-called saddle quantity, $\sigma$; a theorem by Shilnikov (25, 26) states that the dynamics is complex/chaotic for $\sigma > 0$ and simple for negative values (*Materials and Methods*). As shown in Fig. S1, the saddle quantity measured numerically for the TM model is strictly positive and quite large, confirming the presence of Shilnikov chaos.

**Robustness of the Shilnikov Homoclinic Bifurcation.** To examine whether chaos exists under different conditions we varied previously fixed variables, such as $J$, keeping the others fixed. In the 2D parameter space $(I_0, J)$ we found a line of homoclinic bifurcations. Similar lines of homoclinic bifurcations were found by varying other pairs of parameters, such as $(I_0, \tau_D)$, $(I_0, \tau_F)$, $(\tau_D, J)$, $(\tau_F, J)$, and $(\tau_D, \tau_F)$, all of which had large, positive values of the saddle quantity (Fig. S1). In addition, we also studied the effects of synaptic delays on the dynamics of short-term synaptic plasticity (Fig. S2), confirming the robust existence of Shilnikov chaotic dynamics in the TM model.

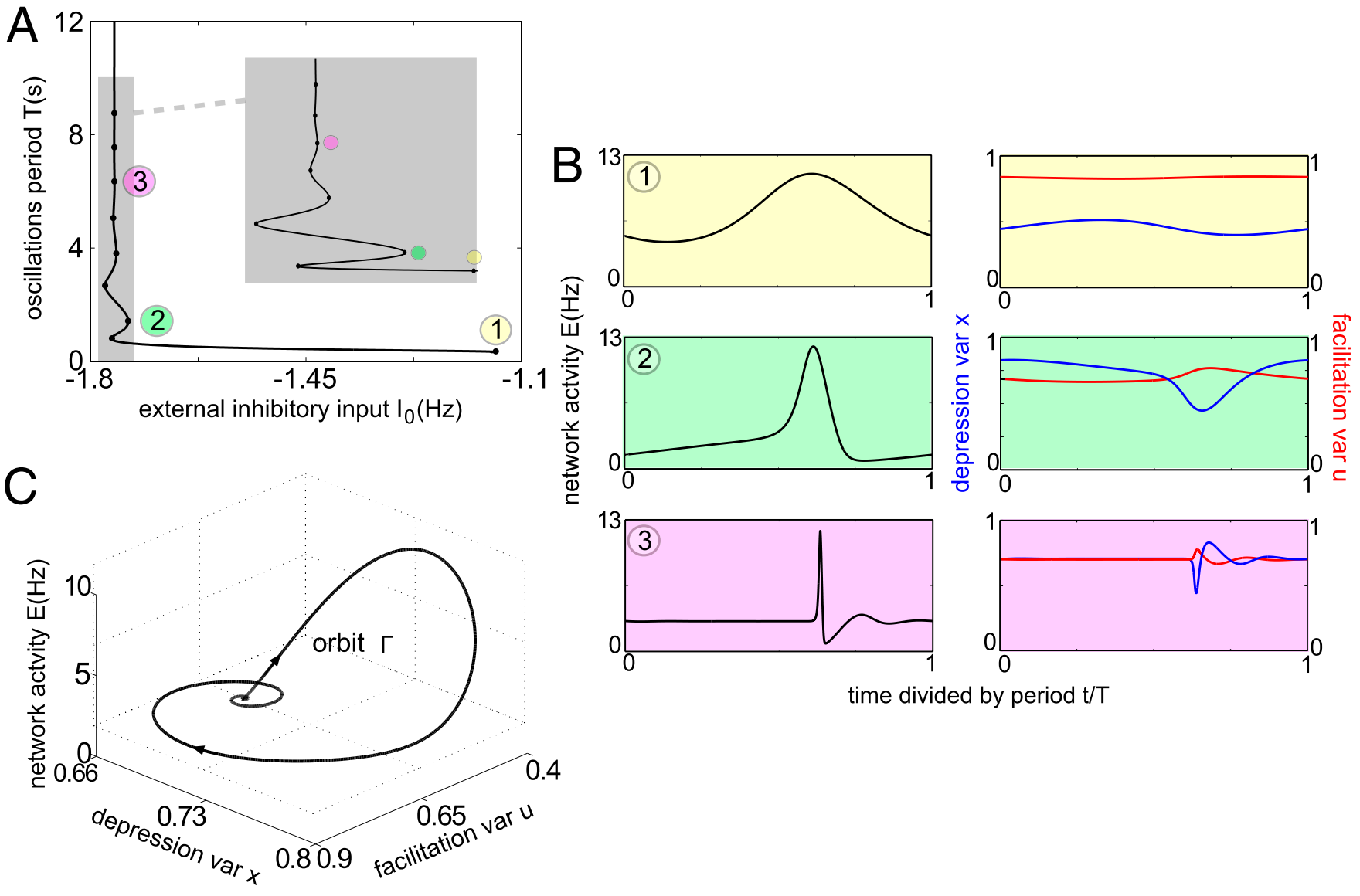

**Fig. 3.** Divergence of the oscillation period as a route to chaos in the STSP dynamics. (*A*) Accumulation of branches of periodic solutions close to a Shilnikov homoclinic (HO) bifurcation. The branch undergoes an infinite number of saddle-node bifurcations (folds); however, these accumulate to a fixed value of $I_0$ corresponding to the Shilnikov homoclinic bifurcation. (*Inset*) A zoom-in of the gray area. (*B*) Time series of the population activity $E$ and synaptic parameters $x$ and $u$ along the branch. Time is measured in units of the period $T$. Far from the homoclinic bifurcation (yellow 1 in *A*) the network activity exhibits a smooth wave with time. Closer to the homoclinic bifurcation, the wave is briefer (green 2). At the homoclinic bifurcation (purple 3), the wave becomes a spike. (*C*) Last periodic orbit Γ along the branch illustrated in *A*, viewed in 3D phase space $(E,x,u)$. The orbit Γ has a large period and is a numerical approximation of the homoclinic orbit (formally with infinite period). The orbit Γ in the plane $(u,x)$ is approximately stable and spirals toward an equilibrium of saddle type in which both stable and unstable directions coexist. The orbit then shoots out in the unstable direction, which approaches the $E$ direction before spiraling back to the saddle equilibria. Simulations are performed for $J = 3.07$; similar behavior was observed for a wide range of parameter values (Fig. S1).

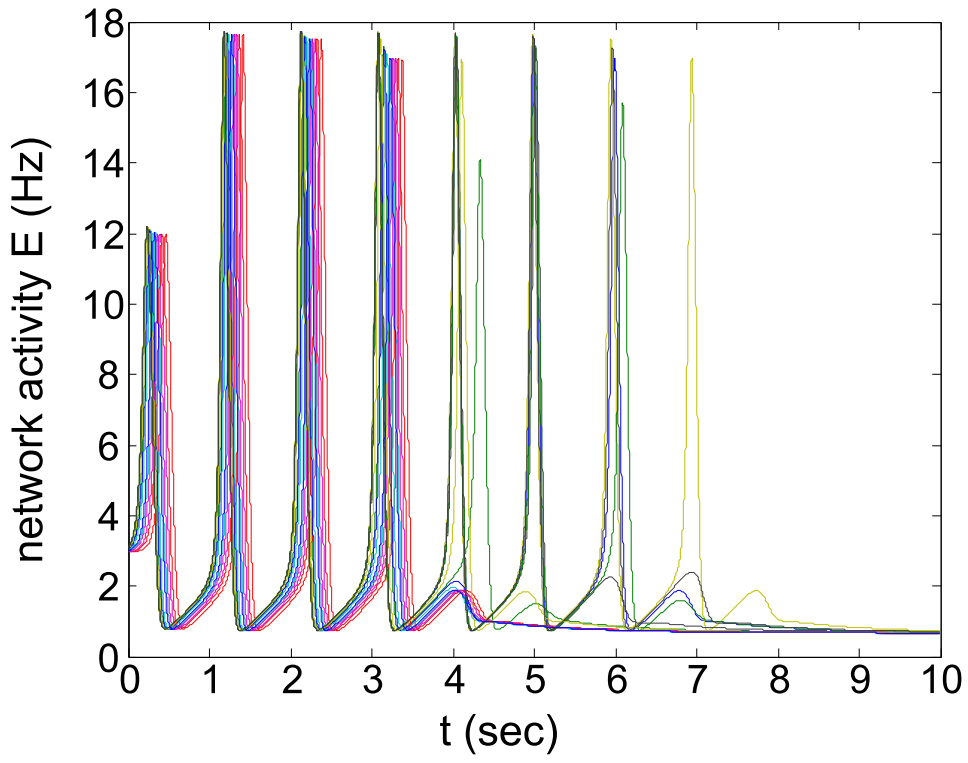

**Fig. 4.** Effects of the Shilnikov chaos in the transient dynamics. Shilnikov chaos differs from other routes to chaos: Although the time series for the TM model near the homoclinic orbit looks regular, the intervals between spikes vary from spike to spike. $E(t)$ is superimposed for trajectories from similar initial conditions $E(t=0)$ (15 different traces in different colors), illustrating the large sensibility to small variations; the trajectories vary in the number of spikes from four to nine before settling down to the stable equilibrium (note also the variable amplitudes of the peaks and the lack of temporal coherence).

**Influence of Noise on the Network Dynamics Near the Shilnikov Homoclinic Bifurcation.** Physiological systems have noise that precludes strict determinism. To study the interplay between stochasticity and the deterministic chaotic dynamics, we have studied a network model proposed in ref. 27 (Fig. 5); the model is such that, in the limit of an infinite number of neurons $N \rightarrow \infty$, it is exactly described by the TM deterministic equations (further details are given in (*SI Text*, section 2). In contrast, for a finite number of neurons, $N$, the network is a noisy version of the TM model with the noise amplitude scaling as $1/\sqrt{N}$. Thus, the standard TM model is the mean field description of the noisy network and the level of noise in the system can be controlled by varying the number of neurons.

We have performed computer simulations of the stochastic network using parameter values close to the homoclinic orbit ($J=3.07, I_0=-1.76$), for which the up-periodic state is at the edge of undergoing a stability change at a deterministic level (Fig. 2). Simulations show that depending on the noise level the network switches between the down state and what seems to be a noisy version of the periodic state (Fig. 5). For intermediate noise intensities (i.e., $N \approx 10^4$), we observed spikes corresponding to the unstable periodic orbits close to the homoclinic point (Fig. 5). As in the deterministic model, these unstable spikes decayed back to the down state and, consequently, were temporally isolated. However, eventually the system jumps to the periodic up state by using the nearby unstable periodic solutions close to the homoclinic orbit as a kind of springboard. Indeed, it did not require much noise to put the system close to the stationary point on the homoclinic orbit, from which the network could jump to nearby unstable periodic orbits. The resulting highly irregular dynamics results from the interplay between deterministic complex dynamics and stochasticity. Further away from the homoclinic bifurcation (i.e., for smaller values of $I_0$), there is no homoclinic orbit between the down- and the up-periodic state and much more noise is required to observe jumps between these two regimes.

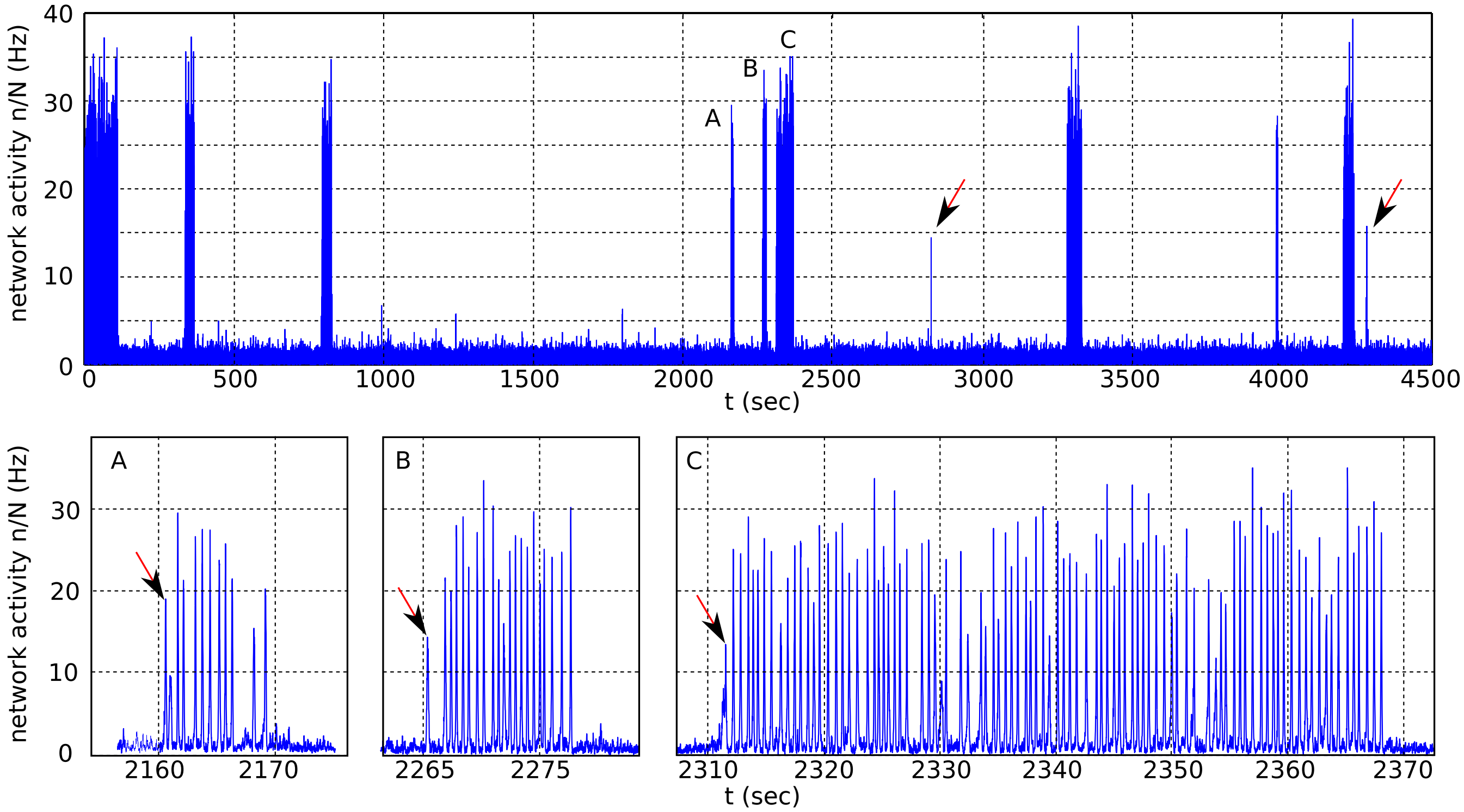

**Fig. 5.** Effects of the Shilnikov chaos in the network dynamics. Simulation of the network activity $n/N$ based on the model introduced in ref. 27, which in its infinite network size limit is described by the TM deterministic equations. We considered a network with $N=17{,}000$ neurons and parameter values $J=3.07$ and $I_0=-1.76$, very close to the homoclinic orbit of the deterministic equations. The initial condition was close to the stationary point on the homoclinic orbit. The standard TM model with these parameters does not have a stable periodic state. (*Upper*) Transitions to noisy transient periodic up states. Some isolated spikes, indicated with an arrow, correspond to the unstable periodic up states near the homoclinic points. (*Lower*) Insets of areas labeled A, B, and C in the upper figure. Observe the high irregularity and complexity of the curves.

## Discussion

**Consequences of the Shilnikov Homoclinic Bifurcation.** We have identified the existence of Shilnikov homoclinic chaos in the simple and parsimonious TM model for the STSP dynamics. Shilnikov proved that if the saddle quantity is positive, infinitely many saddle limit cycles coexist at the bifurcation point: Each of these saddle limit cycles had stable and unstable manifolds, which results in high sensitivity to initial conditions and irregular transient dynamics. In the TM model the saddle quantity is generically positive in all cases: The behavior around the homoclinic bifurcation point is chaotic in the sense of Shilnikov, yielding unpredictable transient patterns with arbitrarily many spikes before settling down to the stable (down-state) equilibrium (Fig. 4).

The striking point to be stressed here is that the presence of a homoclinic bifurcation shapes and organizes the flow of trajectories in a relatively large region around it. We have shown that the Shilnikov bifurcation induces different complex transient responses depending on initial conditions before the system converges to one of the stable attractors. Could this transient dynamics caused by the Shilnikov homoclinic bifurcation mediate or explain certain irregular transients observed in large-scale brain dynamics, such as the erratic and complex transients observed in local field potentials in brain electrophysiology under certain conditions (e.g., ref. 17)? In this context, it is noteworthy that sequences of homoclinic-heteroclinic orbits have been recently proposed to be at the basis of central pattern generators and possibly other high-cognition functions such as memory or motor control in analog, continuous-time, neuronal networks (28); the present study reveals that even isolated homoclinic orbits with strong unstable directions do greatly affect the overall transient dynamics of STSP (and in general the observable network states)—especially when some degree of stochasticity is introduced—and promote transitions between different states, thus fostering complex behavior.

**Simulations of the TM Model on a Network.** Numerical simulations of a network version of the TM model for STSP support the idea that the chaotic manifold can act as a springboard from the down state to the periodic state. The nearby unstable periodic orbits are involved in this switching behavior because they shape the dynamics in an extended region of the phase space. Thus, the network can take advantage of these orbits, through noise, to sample extended regions of the phase space that might not be easily accessible in the deterministic model. This suggests an interesting use of the chaotic manifold as a means to switch between states that are otherwise too far from each other to allow a direct transition through noise.

**Complex Inputs and Balanced States.** We have considered here the effects that constant inhibitory inputs have in the STSP dynamics. However, effects of periodic inputs and other forms of inhibition can be investigated (29). In this case, the usual up state is expected to become oscillatory, and the oscillatory up state to become more complex, with a structure that is likely to depend on the interplay between the intrinsic self-generated oscillations and external ones. It could also potentially lead to much more complex dynamics, such as mixed-mode oscillations and bursting (see ref. 30), by introducing a slow feedback dynamics for the inhibition variable $I(t)$.

It is generally thought that irregular dynamics occurs when the underlying system is in a state in which excitation and inhibition are balanced (31, 32). We have computed the balance ratio in the STSP dynamics, defined as $Ju(t)x(t)E(t)$ divided by $|I_0|$ per time unit along the different trajectories, and verified that all of the active states including fixed points, limit cycles, and chaotic states are generally balanced (i.e., balance ratio of order 1). This is in agreement with ref. 33, which reported that although the excitation–inhibition balance is required for the irregular dynamics, chaotic dynamics is not a necessary ingredient.

**Effects of Noise on Chaotic Dynamics.** Noise—including unreliable synaptic connections, synaptic heterogeneity, irregularity of external inputs, background net activity, network size effects, heterogeneous sparse connectivity, and so on—is ubiquitous in cortical networks and has a profound impact on their dynamics (for a review see ref. 34). Many interesting effects can be found by introducing noise explicitly into the deterministic TM equations or by simulating the TM dynamics upon a (finite) random network (discussed below). For instance, transitions from a down-state fixed point to an up-state fixed point occurred in the presence of stochasticity (11); transitions from either down- or up-state fixed points to periodic up states (two distinct possibilities) were conjectured to exist in ref. 8 and experimentally observed in ref. 35. Also, owing to the presence of noise, up states (but not down states) can exhibit quasi-oscillations, in which the network activity fluctuates in an irregular way—with a nontrivial power spectrum—around its average value (13, 36). The underlying mechanism responsible for such quasi-oscillations is that the corresponding eigenvalues of the underlying up-state fixed point are complex, and hence the transient decay toward it occurs in the form of damped oscillations. Noise eventually kicks the system away from the deterministic fixed point, increasing the amplitude of the damped oscillations, and the system reaches a balanced state in which deterministic relaxation and noise excitation equilibrate, giving rise to a nontrivial spectrum of fluctuations (13, 36). In a similar spirit, we expect that adding noise to the STSP dynamics might lead to highly complex dynamics, especially in the neighborhood of the homoclinic orbit. First, deterministic chaos produces irregular transient spiking behavior (converging to the resting down state as described above). Second, noise kicks the system away from the down state, leading to a complex interplay of chaotic dynamics and stochasticity.

**Self-Organization to the Edge of Chaos.** Recent in vitro experiments revealed that spontaneous up–down transitions occurred predominantly near the limit of stability of the oscillatory phase (37), suggesting that there is some kind of biological self-organizing mechanism tuning cortical networks to the "edge of chaos," a borderline between the active and quiescent phases in parameter space (38–43). This corresponds to the region near the Shilnikov homoclinic bifurcation the TM model unveiled. A feedback loop between excitation and inhibition could be a good candidate to provide for such a mechanism. Future research should explore these issues in the context of more complex network models.

To summarize, we have found the existence of Shilnikov chaos in the parsimonious TM model for short-term synaptic plasticity. Shilnikov chaos may have a profound impact in transient dynamics of large-scale network dynamics, which could be relevant for neural computation. We have shown that the presence of Shilnikov chaos constitutes a possible neural mechanism for transitory variability even in the noiseless neural dynamics and in a stochastic system may serve to facilitate transitions between different stable states.

## Materials and Methods

**Standard TM Model.** We consider the continuous-time version of the TM model (8), as defined by the following set of equations (dot stands for time derivatives):

$$\begin{cases} \tau\dot{E}(t) = -E(t) + g\big(Ju(t)x(t)E(t) + I_0\big) \\ \dot{x}(t) = \tau_D^{-1}\big(1 - x(t)\big) - u(t)E(t)x(t) \\ \dot{u}(t) = UE(t)\big(1 - u(t)\big) - \tau_F^{-1}\big(u(t) - U\big). \end{cases}$$

**Bifurcation Diagram and Numerical Continuation and XPPAUT.** A bifurcation diagram is simply a plot summarizing all possible dynamical states in the parameter space, marking all possible fixed-point steady states, limit cycles, chaotic attractors, and so on, as well as the unstable and saddle points separating their respective domains of attractions. For clarity of visualization, we plotted a one-dimensional projection of the full state space as a function of one parameter: In the case of periodic oscillations only the maximum and the minimum amplitude values are plotted versus the parameter of interest, and the chaotic attractor is represented by a single point.

XPPAUT is a software package that uses continuation techniques to numerically analyze sets of differential equations in an efficient manner (22). By using numerical continuation we tracked the unstable solutions, which cannot be done by with brute-force bifurcation diagrams, in which only stable solutions can be depicted. The code is available at www.math.pitt.edu/~bard/xpp/xpp.html.

**Shilnikov Homoclinic Orbit and Saddle Quantity.** A homoclinic orbit is a trajectory of a dynamical system that joins the stable $W^s$ and unstable $W^u$ manifolds of a saddle-type invariant set with an infinite period. The stable manifold is defined as the set of all trajectories (solutions) that tend to the invariant set in forward time; the unstable manifold is defined as the set of all trajectories that tend to the invariant set in backward time. In this particular case, the invariant dynamical object is a fixed point (here, a saddle-focus equilibrium). The oscillations tend to approximately focus with spiral dynamics; however, the focus possesses only one unstable direction, which has the effect of driving the solution away from this plane before returning again to the oscillatory region (Fig. 3*C*). Thus, in this particular system $W^u$ is one-dimensional, because the equilibrium has one real eigenvalue, $\lambda_1$, and $W^s$ is 2D because of the spiral dynamics, dictated by a pair of complex conjugated eigenvalues $\lambda_{2,3}$ (*SI Text*, section 1 gives the calculation of the eigenvalues). In this way, two eigenvalues $\lambda_{2,3}$ control the stable direction and a third eigenvalue $\lambda_1$ controls the unstable direction.

According to Shilnikov theorem (26), the nature of the overall dynamics of the system near this bifurcation depends on the saddle quantity $\sigma$, defined by $\sigma = \lambda_1 + Re(\lambda_{2,3})$, where $Re$ is the real part of the eigenvalues. In the present case, $\sigma$ can have values that are strictly positive (approximately 15 in Fig. 2, and other positive values in Fig. S1). Consequently, Shilnikov's theorem implies the existence of countably many saddle cycles nearby, which is evidence for chaos. Because there are many saddle-limit cycles dictated by the Shilnikov orbit, an orbit from an initial condition in this region will iterate along these many cycles, thus generating periodic obits with an unpredictable number of spikes, *N*, at irregular intervals.

The classic Shilnikov theorem implies that the unstable direction is associated with the real eigenvalue and that the stable one is associated with the complex conjugate eigenvalues. In case of having directions opposite to the classic case (i.e., stable for the real eigenvalue and unstable for the complex conjugate eigenvalues), the positivity of the saddle quantity does not imply chaos. In any case, the absolute value of the real eigenvalue has to be bigger than the absolute value of the real part of the complex conjugate eigenvalues, which is equivalent to the positivity of the saddle quantity in the classic Shilnikov scenario (i.e., the one we report here), and to its negativity in the scenario with opposite directions.

**Two-Parameter Continuation for Computation of the Chaotic Region.** To estimate the volume in parameter space containing chaotic dynamics, we used a numerical strategy based on an extended boundary-value problem, pseudo-arc-length continuation (22). We computed a two-parameter family of approximate Shilnikov homoclinic orbits (similar to that shown in Fig. 3*C*) together with its associated saddle-focus equilibrium. More specifically, we use the module HOMCONT, part of the software package AUTO (44), which performs these types of computations, which were not possible with XPPAUT. In this way, we were able to compute the saddle quantity $\sigma$ associated with the saddle-focus equilibrium for every couple of parameter values along the computed family. For a saddle-focus equilibrium of a 3D vector field, this quantity is defined as the sum of its real eigenvalue and the real part of its complex conjugated eigenvalues. A major theoretical result states that if $\sigma$ is strictly positive, then the chaotic dynamics extends to a neighborhood around the parameter value of the Shilnikov homoclinic bifurcation (25).

**ACKNOWLEDGMENTS.** J.M.C. is supported by Ikerbasque, the Basque Foundation for Science. This work was supported by Junta de Andalucia Grant P09-FQM-4682 (to J.M.C. and M.A.M.), the Howard Hughes Medical Institute, and Office of Naval Research Grant N000141210299 (to R.V. and T.J.S.).